\documentclass[useAMS,usenatbib]{mn2e}
\usepackage{amsmath}
\DeclareTextSymbol{\degre}{OT1}{23}
\usepackage{graphicx}

\title[Masses of SB2 components]{Masses of the components of SB2 binaries observed with {\it Gaia}. II.
Masses derived from
{\it PIONIER} interferometric observations for {\it Gaia} validation
\footnotemark[1]\thanks{based on observations performed
at the Observatoire de Haute--Provence (CNRS), France}
\footnotemark[2]\thanks{based on observations performed with the {\it HERMES} spectrograph
at the Roque de los Muchachos Observatory}
\footnotemark[3]\thanks{based on data obtained with the ESO Very Large Telescope under
programme 094.D-0624, 094.C-0884, 094.C-0175 and 094.C-0397.}}
\author[J.-L. Halbwachs et al.]{J.-L. Halbwachs$^{1}$\thanks{E-mail:
jean-louis.halbwachs@astro.unistra.fr},
H.M.J. Boffin$^{2}$, J.-B. Le Bouquin$^{3}$, F. Kiefer$^{4}$, B. Famaey$^{1}$,
  \newauthor
J.-B. Salomon$^{1}$, F. Arenou$^{5}$, D. Pourbaix$^{6,7}$, F. Anthonioz$^{3}$, R. Grellmann$^{8}$,
  \newauthor
S. Guieu$^{3}$, H. Sana$^{9}$, P. Guillout$^{1}$, A. Jorissen$^{7}$, Y. Lebreton$^{5,10}$, T. Mazeh$^{4}$,
  \newauthor
L. Tal-Or$^{4,11}$ and A. Nebot G\'omez-Mor\'an$^{1}$\\
$^{1}$Observatoire astronomique de Strasbourg, Universit\'e de Strasbourg, CNRS, UMR 7550,
11 rue de l'Universit\'{e}, 67000 Strasbourg, France\\
$^{2}$ESO, Av. Alonso de Cordova 3107, 19001 Casilla, Santiago 19, Chile\\
$^{3}$UJF-Grenoble 1/CNRS-INSU, UMR 5274, Institut de Plan\'etologie et d'Astrophysique de Grenoble (IPAG), 38041 Grenoble, France\\
$^{4}$School of Physics and Astronomy, Tel Aviv University, Tel Aviv 69978, Israel\\
$^{5}$GEPI, Observatoire de Paris, CNRS, Universit\'e Paris Diderot, 5 place Jules Janssen, 92195, Meudon, France\\
$^{6}$F.R.S.-FNRS, Rue d'Egmont 5, 1000 Bruxelles, Belgium\\
$^{7}$Institut d'Astronomie et d'Astrophysique, Universit\'{e} Libre de Bruxelles, boulevard du Triomphe, 1050 Bruxelles, Belgium\\
$^{8}$Physikalisches Institut der Universit\"at zu K\"oln, Z\"ulpicher Strasse 77, D-50937 K\"oln, Germany\\
$^{9}$Institute of Astronomy, KU Leuven, Celestijnenlaan 200D, 3001 Leuven, Belgium\\
$^{10}$Institut de Physique de Rennes, Universit\'e de Rennes 1, CNRS UMR 6251, 35042 Rennes, France\\
$^{11}$Institut f\"ur Astrophysik (IAG), Friedrich-Hund-Platz 1, D-37077 G\"ottingen, Germany}

\begin{document}

\date{Accepted 2015 October 23.  Received 2015 October 20; in original form 2015 September 30}

\pagerange{\pageref{firstpage}--\pageref{lastpage}} \pubyear{2015}

\maketitle

\label{firstpage}

\begin{abstract}
In anticipation of the {\it Gaia} astrometric mission, a sample of spectroscopic binaries is being observed
since 2010 with the {\it Sophie} spectrograph at the Haute--Provence Observatory. Our aim is to derive the orbital elements of double-lined
spectroscopic binaries (SB2s) with an accuracy sufficient to finally obtain the masses of the components
with relative errors as small as 1~\% when combined with {\it Gaia} astrometric measurements.
In order to validate the masses derived from {\it Gaia}, interferometric observations are obtained for three SB2s in our sample with F-K components: HIP 14157, HIP 20601 and HIP 117186. 
The masses of the six stellar components are derived. 
Due to its edge-on orientation, HIP 14157 is probably an eclipsing binary. 
We note that almost all the derived masses are a few percent larger than the expectations from the standard spectral-type-mass calibration and mass-luminosity relation.
Our calculation also leads to accurate
parallaxes for the three binaries, and the {\it Hipparcos} parallaxes are confirmed.

\end{abstract}


%=======================================================================================================================

\section{Introduction}

In a previous paper \citep{Halb2014}, we presented the selection of a sample of double-lined spectroscopic binaries
(SB2) for which it will be possible to derive accurately the masses of the components when the astrometric measurements of
the {\it Gaia} satellite will be delivered. Our aim is to obtain high-precision radial velocity (RV) measurements in
order to derive the minimum masses of the components, ${\cal M}_1 \sin^3 i$ and ${\cal M}_2 \sin^3 i$, where ${\cal M}_1$
and ${\cal M}_2$ are the masses and $i$ is the inclination of the orbital plane on the plane of the sky. The {\it Gaia} 
astrometric measurements of the
photocentre of these systems will lead to the derivation of astrometric orbits, including $i$, and therefore to ${\cal M}_1$ and ${\cal M}_2$.
The RV measurements are obtained through two different programs: a programme of about 70 SB2 is carried on with the
T193 telescope of the Haute-Provence observatory ({\it OHP}) with the {\it Sophie} spectrograph, and a separate programme of seven SB2 is using the
Mercator 1.2m--telescope at the Roque de los Muchachos Observatory ({\it RMO}), with the {\it HERMES} spectrograph \citep{Raskin}.

Despite the high quality that is expected for the {\it Gaia}
measurements, we know from the reduction of the {\it Hipparcos} satellite \citep{ESA97,vanLeeuwen2007} that large space
astrometric surveys may be prone to systematic errors. The discussion about the reliability of the {\it Hipparcos} results is still
not closed: recently, \citet{Fekel15} found an important discrepancy between the parallax he obtained for the system
HD 207651, and the parallax given in the {\it Hipparcos 2} catalogue \citep{vanLeeuwen2007}, even when the orbital motion
is taken into account.

The large variations affecting the basic angle of {\it Gaia} will make the verification of the {\it Gaia} measurements even more necessary
\citep{Mora2015}. Regarding double stars,
a good orbit determination with {\it Gaia} depends on the measurements
at various epochs, and ultimately rely on the application of the Point Spread Function (PSF) calibration, considering the object as a single star.
Binaries may then be sensitive to small differences between the position given by the PSF and the position of the actual photocentre.
Therefore, an independent derivation of the masses of some stars of our sample is welcome in order to validate our future results.

For these reasons, we obtained interferometric measurements with the {\it PIONIER} instrument at the Very Large Telescope Interferometer ({\it VLTI}).
After one semester, three SB2s were observed over nearly half of their period. Two of these systems are from the {\it OHP}
programme, and the other is from the {\it RMO}  programme. HIP~14157 is a couple of chromospherically active K-dwarf stars which was noticed
by \citet{Fekel04} for their large minimum masses. HIP~20601 is a Hyades binary star observed by \citet{Griffin85}, and that seems also
to have components with masses above expectations. HIP~117186 is an early F-type dwarf from the sample of \citet{Nordstrom1997}.

The obtention of the interferometric observations is described in section~\ref{sec:PIONIER}. The RV data and the calculation of the
SB2 orbital elements are in section~\ref{sec:SB2}. The masses of the six components are derived hereafter, in section~\ref{sec:masses}. 
Our results are used to derive the mass-luminosity relation in the infrared $H$ band,
in section~\ref{sec:masse-lum}, and also to verify the {\it Hipparcos} parallaxes, in section~\ref{sec:varpi}. We conclude in 
section~\ref{sec:conclusion}.

%_________________________________________________________________________________________________________________

\section{The interferometric observations}
\label{sec:PIONIER}

\subsection{The observations}

We observed our three systems with the four 1.8-m Auxiliary Telescopes of {\it ESO} Very Large Telescope Interferometer, using the {\it PIONIER} instrument \citep{Berger10, Pionier11} in the $H$-band on several nights:
6--8 October 2014, 17--20 October 2014, 31 October--1 November 2014, 15--17 November 2014, 4--5 December 2014, 27--28 January 2015, and 5--6 February 2015. For all targets observed in 2014, we used the prism in low resolution ({\it SMALL}) which provides a spectral resolving power R$\sim$15, the fringes being sampled over three spectral channels. 
Mid-December 2014, the detector of {\it PIONIER} was changed to the new {\it RAPID} detector, and therefore observations obtained in 2015 were obtained with the new observing mode, sampling the fringes over six spectral channels.
The large {\it VLTI} configuration A1--G1--K0--J3 was used -- except on the nights of 15--17 Nov. 2014 when
A1--G1--I1--K0 was used and on 5--6 Feb. 2015 when H0--I1--D0--G1 was used -- leading to baselines of 41 (H0-I1), 47 (G1-I1, H1-I1 and K0-I1), 56.8 (K0-J3), 64 (D0-H0), 71.7 (D0-G1 and H0-G1), 80 (A1-G1), 82.5 (D0-I1), 91 (G1-K0), 107 (A1-I1), 129 (A1-K0), 132.4 (G1-J3), and 140 metres (A1-J3).

Data reduction and calibration were done in the usual way with the {\tt  pndrs} package presented by \citet{Pionier11}.
Each pointing provides six visibilities and four closure phases dispersed over the few spectral channels across the {\it H}-band.

These interferometric observations were adjusted by a simple binary model. The diameters of
the individual components, all smaller than $0.21$\,mas, are unresolved by our instrumental setup.
The free parameters are thus the separation, the position angle of the secondary with respect to the primary, and the flux ratio.
The different epochs were mostly adjusted independently. For few epochs, corresponding to small separations or incomplete dataset,
the flux ratio was imposed following the results obtained at other epochs.

Fitting a binary model to interferometric observations is non-linear and non-convex\footnote{Broadly speaking, a convex optimisation means that there is only one optimal solution, which is therefore globally optimal, i.e. a local minimum is a global one. This is not the case when adjusting a binary model to interferometric observations, as we need to cover the full range of parameters to determine the position of the global minimum.}.
We used a classical gridding approach to overcome these issues and to find the deepest minimum.
We then used a Levenberg-Marquardt algorithm to determine the best-fit parameters and their covariance matrix.
The astrometric error ellipsoid is the on-sky representation of this covariance matrix.

\begin{figure*}
\includegraphics[clip=,height=14.8 cm]{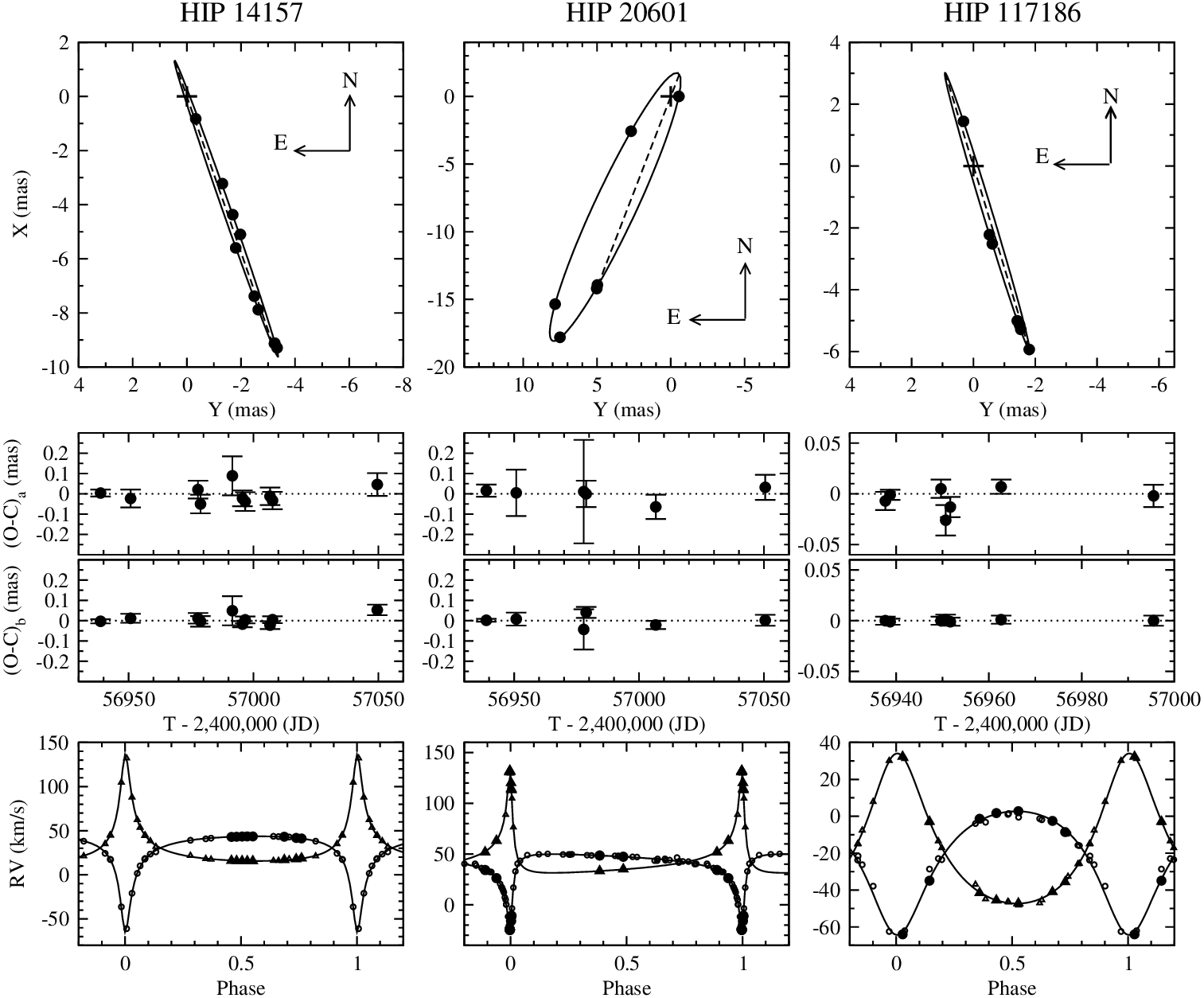}
 \caption{The combined orbits of the three SB2 observed with PIONIER.
Upper row: the visual orbits; the node line is in dashes, while the position of the primary is indicated by the cross. Second row:
the residuals along the semi-major axis of the error ellipsoid. Third row:
the residuals along the semi-minor axis of the error ellipsoid. Last row:
the spectroscopic orbits; the circles refer to the primary component, and the
triangle to the secondary; the large filled symbols refer to the new RV measurements
obtained with {\it HERMES} or {\it Sophie}. For each SB2, the RVs are shifted to the
same zero point.
}
\label{fig:orbites}
\end{figure*}

Our results are summarised in 
Table~\ref{tab:mesPIO}. The positions of the secondary component with respect to the primary are plotted on
Fig.~\ref{fig:orbites}.

\begin{table}
 \caption{The interferometric measurements. The column $f_2/f_1$ gives the flux ratio between the two components in the infrared $H$ band.  
$\rho$ is the
separation between the components, and $\theta$ is the position angle of the secondary with
respect to the primary component. $\sigma_{max}$ and  $\sigma_{min}$ are the semi-major axis and
the semi-minor axis of the error ellipsoid, respectively, corrected as indicated in 
section~\ref{sec:correction}. The last column
is the position angle of the semi-major axis of the  error ellipsoid.
The flux ratios flagged with asterisks were fixed in the fitting procedure.
}
\label{tab:mesPIO}
\begin{tabular}{@{}lcrrllr}
\hline
\multicolumn{7}{c}{HIP 14157} \\
 &&&&&&\\
MJD & $f_2/f_1$ & $\rho$ & $\theta$ & $\sigma_{max}$ &  $\sigma_{min}$ & PA   \\
          &           & mas    &   deg    &    mas    &    mas     & deg  \\
\hline
56938.290 & 0.663 &  5.88 & $-$162.05 &  0.017 &  0.010 &     132 \\
56950.291 & 0.683 &  3.48 & $-$157.77 &  0.044 &  0.022 &     118 \\
56977.272 & 0.676 &  8.31 & $-$161.43 &  0.044 &  0.026 &     155 \\
56978.285 & 0.691 &  7.79 & $-$161.28 &  0.046 &  0.026 &     159 \\
56991.054 & 0.670$^*$ &  0.89 & $-$158.44 &  0.096 &  0.072 &   110 \\
56995.109 & 0.688 &  4.69 & $-$158.73 &  0.039 &  0.015 &     150 \\
56996.165 & 0.676 &  5.47 & $-$158.75 &  0.046 &  0.017 &     149 \\
57006.115 & 0.662 &  9.68 & $-$160.33 &  0.043 &  0.019 &     153 \\
57007.140 & 0.664 &  9.87 & $-$160.19 &  0.043 &  0.017 &     143 \\
57049.045 & 0.666 &  9.69 & $-$160.41 &  0.056 &  0.026 &     129 \\
\hline
\multicolumn{7}{c}{HIP 20601} \\
 &&&&&&\\
 MJD & $f_2/f_1$ & $\rho$ & $\theta$ & $\sigma_{max}$ &  $\sigma_{min}$ & PA   \\
          &           & mas    &   deg    &    mas    &    mas     & deg  \\
\hline
56938.367 & 0.402 &  3.72 & +133.73 &  0.030 &  0.0067 &     139 \\
56950.348 & 0.400$^*$ &  0.55 &  $-$89.55 &  0.11  &  0.032 &     150 \\
56977.353 & 0.400$^*$ & 14.79 & +160.44 &  0.25  &  0.099 &     148 \\
56978.321 & 0.400$^*$ & 15.06 & +160.61 &  0.065 &  0.027 &     149 \\
57006.187 & 0.402 & 19.32 & +157.20 &  0.060 &  0.020 &     157 \\
57050.051 & 0.387 & 17.24 & +153.02 &  0.062 &  0.027 &     162 \\
\hline
\multicolumn{7}{c}{HIP 117186} \\
 &&&&&&\\
MJD & $f_2/f_1$ & $\rho$ & $\theta$ & $\sigma_{max}$ &  $\sigma_{min}$ & PA   \\
          &           & mas    &   deg    &    mas    &    mas     & deg  \\
\hline
56937.179 & 0.440 &  2.28 & $-$166.93 &  0.0090 &  0.0036 &       8 \\
56938.210 & 0.440 &  2.59 & $-$166.40 &  0.0045 &  0.0027 &     163 \\
56949.162 & 0.434 &  5.20 & $-$164.20 &  0.0086 &  0.0036 &     139 \\
56950.187 & 0.437 &  5.35 & $-$163.79 &  0.015 &  0.0050 &     134 \\
56951.189 & 0.454 &  5.50 & $-$163.84 &  0.0099 &  0.0036 &     141 \\
56962.137 & 0.430 &  6.20 & $-$163.09 &  0.0072 &  0.0036 &     141 \\
56995.017 & 0.446 &  1.48 &  +12.55 &  0.011 &  0.0045 &     168 \\
\hline
\end{tabular}

\end{table}

\subsection{Verification and correction of the uncertainties}
\label{sec:correction}

Reliable uncertainties are needed to derive masses from apparent positions and from RV measurements.
This point is especially important hereafter, since several parameters in the common solution leading to the
masses are coming as well from the interferometric observations as from the spectroscopic ones. This applies
to the period $P$, the eccentricity $e$, the epoch of the periastron $T_0$ and the periastron longitude $\omega$.
Overestimating the uncertainties of the interferometric observations would lead to underestimate the weights
of these observations and to exaggerate the contribution of the RV in the derivation of these terms, and
vice versa.

To verify the uncertainties, the ``visual'' orbit of the star is derived. Computing a visual orbit consists
in searching seven unknowns: the period, $P$, the eccentricity, $e$, the epoch of the periastron, $T_0$,
and the Thiele-Innes elements $A$, $B$, $F$, $G$. Therefore, a minimum
of eight observations is needed to derive these terms and to estimate their errors. Since a relative position
is a two-dimensional observation, this corresponds to four interferometric observations. For the least observed star, which is
HIP~20601, we have six interferometric observations, resulting in five degrees of freedom in the derivation of the orbit. This is
sufficient to allow the verification of the error estimations.

We derive the orbit of each binary, and we calculate $F_2$, the estimator of the goodness-of-fit
defined in \citet{Kendall}:

\begin{equation}\label{def:F2}
F_2= \left( \frac{9\nu}{2} \right)^{1/2} \left[ \left( \frac{\chi^2}{\nu} \right)^{1/3}+{\frac{2}{9 \nu}}-1 \right]
\end{equation}

where $\nu$ is the number of degrees of freedom and $\chi^2$ is the weighted
sum of the squares of the differences between the predicted and the observed
values, normalised with respect to their uncertainties. When the predicted values
are obtained through a linear model, $F_2$ follows the normal distribution
${\cal N}(0,1)$. When non--linear models are used, but when the errors are small
in comparison to the measurements, as hereafter, the model is approximately linear around the
solution, and $F_2$ follows also ${\cal N}(0,1)$. We also used simulations to verify
that this property is true even when the number of degrees of freedom is as small as five.

It appears that, when the uncertainties provided by the {\it PIONIER} reduction are taken into account, $F_2$ is systematically
negative. This indicates that these uncertainties are overestimated. In order to keep the relative weights of the
observations, the uncertainties of the positions of any star are divided by the same coefficient, in order to
have $F_2=0$. The uncertainties $\sigma_{min}$ and $\sigma_{max}$ in Table~\ref{tab:mesPIO} are thus obtained. It is
worth noticing that they are similar to the errors expected for {\it Gaia} \citep{Perryman05,Eyer15}.

%_________________________________________________________________________________________________________________

\section{The radial velocities and the SB2 orbits}
\label{sec:SB2}

\subsection{Existing RV measurements}
\label{sec:oldRV}

Radial velocity measurements are obtained from the SB9 Catalogue \citep{SB9}, which is regularly updated and accessible on-line\footnote{http://sb9.astro.ulb.ac.be/}. Primarily, the spectroscopic orbits of these stars were derived assigning weights to the measurements
of the primary and of the secondary component.  Since our purpose is to derive masses not only from these measurements, but also from
new RV measurements and from interferometric observations, it is necessary to convert these weights to reliable uncertainties.

To evaluate the uncertainty of the RV of any component, the single-lined orbit of the star is derived. The weights of the RV measurements
are transformed to uncertainties in order to get a SB1 orbit with $F_2=0$.
It is worth noticing
that this method is similar in its principle to the usual approach consisting in taking for the
uncertainty the standard deviation of the residuals of the RV. However, it is much more reliable and it leads to
uncertainties significantly larger.

When the RV uncertainties are obtained for both components, the SB2 orbit is derived, and $F_2$ is
calculated again. A correction coefficient is applied to have at the end an SB2 orbit again with
$F_2=0$. This method leads to the following results:

\begin{itemize}

\item
For HIP~14157, the uncertainty of the RV measured by \citet{Fekel04} is 0.582 and
0.677 km s$^{-1}$ for the primary and for the secondary component, respectively.

\item
The standard procedure requires an adaptation for the treatment of the RV measurements provided by
\citet{Griffin85} for HIP~20601. No weights are indicated by the authors, but 7 of the
63 measurements of the primary component are flagged as uncertain, and the secondary
component received only 4 RV measurements. We assign 0.843~km s$^{-1}$ to the uncertainty of the 56 
``not uncertain'' RV measurements, in order to get an SB1 orbit with $F_2=0$. The uncertainty
of the ``uncertain'' primary RV is then 1.504~km s$^{-1}$, in order to still have $F_2=0$ for the SB1
orbit of the primary component. The uncertainty of the RV of the secondary component is then
3.000~km s$^{-1}$, in order to have $F_2=0$ for the SB2 orbit of the binary.

\item
For HIP~117186, the uncertainty of the RV measured by \citet{Nordstrom1997} is 1.837 and
1.487 km s$^{-1}$ for the primary and for the secondary component, respectively.

\end{itemize}

The sets of RV of the three stars are further completed with high-accuracy measurements
recently obtained.

%-----------------------------------------------------------------------------------------------------------------
\subsection{New RV measurements}

\subsubsection{Radial velocities from the {\it HERMES} spectrograph}
\label{sec:RV_HERMES}

HIP~14157 received 8 RV measurements between January 2015 and August 2015. The fibre-fed {\it HERMES} spectrograph
covers the whole wavelength range from 380 to 900~nm at a resolving power of $\sim 86,000$. A Python-based pipeline 
extracts a wavelength-calibrated and a cosmic-ray cleaned spectrum. A restricted region, covering the range
478.11 -- 653.56~nm (orders 55 -- 74) was used to derive a cross-correlation function (CCF) with a spectral mask constructed
from an Arcturus spectrum and containing  2103 useful spectral lines. A
spectrum with a signal-to-noise ratio of 15 is usually sufficient to
obtain a cross-correlation function (CCF) with a well-pronounced
maximum.

Radial velocities are determined from a Gaussian fit to the core of the CCF 
with an internal precision of a few m~s$^{-1}$. The most important external source of error is the varying atmospheric 
pressure in the spectrograph room \citep[see Fig.~9 of][]{Raskin}, which is largely eliminated by the arc spectra taken 
for wavelength calibration. The long-term stability (years) of the resulting radial velocities is checked with RV standard 
stars from \citet{Udry99}. 
Their standard-deviation distribution peaks at $\sigma_{RV}=55$~m~s$^{-1}$, which we adopt as the typical radial-velocity 
uncertainty for such relatively bright single stars. The RV standard stars have also been used to tie the {\it HERMES} RVs to the IAU standard system.

For a SB2 system, the reduction process leads to estimations of the uncertainties of the RV which are obviously underestimated, due to the
pollution of the spectrum of each component by that of the other one.
This appears clearly when the orbital elements are derived from the 8 {\it HERMES} observations of HIP 14157: the goodness-of-fit of the SB2 solution
is as large as $F_2=17.3$, and the standard deviations of the residuals are 0.138 and 0.219~km~s$^{-1}$, respectively. We assume
then that the ratio of the true uncertainties is $\sigma_{RV2}/\sigma_{RV1}=1.59$. The uncertainties are then
increased, by adding quadratically a noise depending on the component, until we have $F_2=0$. This condition is fulfilled
with the noises 0.187 and 0.298~km s$^{-1}$, respectively. This method is a bit different from the one applied above to
derive the uncertainties of the previously published
measurements, since the correction of the uncertainties is not done separately for each component. However, it is 
more suitable for an SB2 with few observations. The RVs of the components of HIP~14157 and the uncertainties thus obtained
are listed in Table~\ref{tab:VR}.

%-----------------------------------------------------------------------------------------------------------------
\subsubsection{Derivation of RV from {\it Sophie} spectra with {\it TODMOR}}

HIP~20601 and HIP~117186
received 12 and 7 {\it Sophie} spectra, respectively, between December 2010 and December 2014.
The RV of the components are derived using the two-dimensional correlation algorithm
{\it TODCOR} \citep{zucker94,zucker04}, as explained hereafter.

The {\it TODCOR} algorithm calculates the cross-correlation of an SB2 spectrum and two best-matching stellar atmosphere models, one for each component of the observed binary system. This two-dimensional cross-correlation function (2D-CCF) is maximized at the radial velocities of both components. The multi-order version of {\it TODCOR}, named {\it TODMOR}~\citep{zucker04}, determines the radial velocities of both components from the gathering of the 2D-CCF obtained from each order of the spectrum.

All {\it Sophie} multi-orders spectra are deblazed, then pseudo-continuum normalised using a 
$p$-percentile filter~\citep{Hodg1985}. 
The percentile $p$=$0.5$ selects the median among all flux values contained within the filter's window; any $p$$>$$0.5$ selects flux 
with value larger than the median. The percentile $p$ and the width $w$ of the filtering window are chosen so that the resulting 
normalised spectra are as flat as possible, while not altering the depth and shape of any lines.
These constraints led us to chose for both targets $w\!\!\sim$1,300 pixels ($\sim$33\AA); the width of one {\it Sophie} order is about $4,000$ pixels ($\sim$100\AA); 33\AA ~is about twice the full width at half-maximum for a Balmer line of early-type stars and for Ca II lines of late-type stars. The value of $p$ was determined independently for each order by maximising the two dimensional cross-correlation.

We determined for the two components of each binary best-matching atmospheric models from the {\it PHOENIX} library~\citep{Huss2013}. For consistency, we also applied a $p$-percentile filter on the spectra 
of the models, with $p$=$0.99$ and the same window's width as for HIP~117186 and HIP~20601 spectra, 
of $\sim$1,300 pixels. On those spectra for which both peaks of the SB2 
components are well separated, we optimized the two-dimensional cross-correlation function varying 
the effective temperature, the stellar rotation's $v\sin i_r$, the metallicity and the surface gravity's 
$\log g$ of both components. 

The grid for optimisation is defined with $T_\text{eff}$ extending from 3,000 to 6,900~K with steps of 100~K and from 7,000 to 11,000 with steps of 200~K; 
$\log(g)$ extending from 2 to 6 dex with steps of 0.5 and linearly interpolated from 4 to 5.5 with steps of 0.1; [Fe/H] extending from $-1.5$ to 1 dex with steps of 0.5 and linearly interpolated from $-1$ to 0.5 with steps of 0.1; $v \sin i_r$ extending continuously from 
0 to 200~km.s$^{-1}$; and the flux ratio, $\alpha$, extending continuously from 0 to 1. The secondary $\log g$ was fixed for each set of test parameters with respect to primary and secondary effective temperatures, primary $\log g$ and estimated mass ratio $q$=$M_2/M_1$, using the following relation

\begin{align}
\log g_2 - \log g_1 & = \log q + 2 \log\frac{R_2}{R_1} \nonumber \\
                    & = \log q + \log \alpha - \log \frac{B(T_{\text{eff},2}, \lambda_{med,20})}{B(T_{\text{eff},1},\lambda_{med,20})} 
\end{align}

where $B(T_\text{eff})$ is the value of the blackbody flux for effective temperature $T_\text{eff}$ at the median wavelength $\lambda_{med,20}$ of \emph{Sophie}'s median order n$^\circ$19 over 39.   

The derived values of the stellar parameters are given for both components in Table~\ref{tab:stellpar}. 
For HIP~117186 only one spectrum had large enough separation between the primary and secondary peaks, namely the spectrum observed at periastron passage. For HIP~20601, all spectra gave very consistent results, up to 5K and 100K in effective temperature for the primary and secondary respectively, 0.03 and 0.4 in $\log g$, 0.2 and 0.6 km\,s$^{-1}$ in v\,$\sin i_r$, 0.01 in metallicity, and 0.01 in flux ratio. However, the individual uncertainties are much larger than these scatter values, so we give here the average individual uncertainties divided by the square-root of the number of spectra used for deriving the parameters, namely $\sqrt{N_{\rm spec}}$.

The individual uncertainties were determined by defining as lower and upper bounds the values of the parameters at maximum of the CCF minus the estimated level of the noise in the CCF. This level is given by $\sigma_{\rm CCF}$$\sim $$\sqrt{\langle 1/{\rm SNR}^2\rangle}/\Vert F \Vert$, the order-average noise-to-signal ratio in the normalised spectrum $F$ divided by its norm over all orders

\begin{align}
\langle 1/{\rm SNR}^2\rangle &= \frac{1}{N_{orders}}\times\sum_{i \in {\rm orders}} \left(\frac{1}{{\rm SNR}_i}\right)^2 \\
\Vert F \Vert & =\sqrt{\sum_{i \in {\rm orders}}\sum_{p \in {\rm pixels}} F_{i,p}^2}
\end{align}

\begin{table}
 \centering
 \begin{minipage}{70mm}
 \caption{The stellar parameters determined by optimisation of the two-dimensional cross-correlation function obtained with {\it TODMOR}. At the bottom, $N_{\rm spec}$ is the number of spectra used to derive the parameters values.}

\label{tab:stellpar}
\begin{tabular}{@{}lll}
\hline %--------------------------------------------------------------
Parameters                                     &   HIP 20601        &  HIP 117186 \\
\hline %--------------------------------------------------------------
$ T_{\rm eff,1} {\rm [K]}$                     &    5600$\pm$90     & 6580$\pm$230    \\
$ \log g_{\rm 1} {\rm [dex]}$                  &    4.43$\pm$0.26   & 3.8$\pm$0.6    \\
$ v_1$ $\sin i_{r \; \rm 1} [{\rm km s}^{-1}]$ &    4.6$\pm$1.0     & 43$\pm$13    \\
 $T_{\rm eff,2} {\rm [K]}$                     &    4550$\pm$550    & 6550$\pm$490    \\
 $\log g_{\rm 2} {\rm [dex]}$                  &    4.76$\pm$0.92   & 4.32$\pm$0.82     \\
 $v_2$ $\sin i_{r \; \rm 2} [{\rm km s}^{-1}]$ &    $<$2              & 13$\pm$11     \\
 ${\rm m/H} {\rm [dex]}$                       &   -0.38$\pm$0.10   & -0.35$\pm$0.24  \\
 $\alpha$ [flux ratio]                         &    0.106$\pm$0.006 &  0.374$\pm$0.151  \\
&& \\ %\hline %---------------------------------------------------------------
$N_{\rm spec}$                                 &  4                 &   1               \\
\hline %---------------------------------------------------------------
\end{tabular}
\end{minipage}
\end{table}

Finally, we applied {\it TODCOR} to all multi-order spectra of each target and determined the radial velocities of both components. We discarded several orders of the spectra of HIP~20601 and HIP~117186, which were strongly affected by telluric lines (orders n$^\circ$31, 34, 36 and 39). At a given exposure, for each of the selected orders, we calculated a two dimensional cross-correlation function, from which we derived the maximising values of radial velocities for the primary and the secondary.

For each target, there are systematic order-to-order variations of the radial velocity measurements, different for each SB2 component. These systematics come from signal-to-noise, number of atomic lines available, and discrepancies of the models with the real spectrum; they are specific to each component and each order. 
To estimate them, we first calculated for each exposure the residuals of the velocities derived for all individual order around the median velocity; then we considered the residuals obtained for all exposures at an individual order, and calculated the systematic shift for this order as the median of the residuals. We estimated as well a measurement error for each order from the scatter of its residuals about the systematics. 

Then, we determined at all epochs the radial velocities of the primary and the secondary and their uncertainties from the weighted average and the square-root of the weighted variance of the corrected order-by-order velocities. \\

\begin{table}
 \centering
 \begin{minipage}{70mm}
 \caption{The new RVs obtained from {\it HERMES} (HIP~14157) or from {\it Sophie} (HIP~20601 and HIP~117186).
The uncertainties are revised as explained in the text.}
\label{tab:VR}
\begin{tabular}{@{}lrrrr}
\hline
\multicolumn{5}{c}{HIP 14157} \\
 &&&&\\
 BJD        & $RV_1$      & $\sigma_{RV\;1}$ & $RV_2$       & $\sigma_{RV\; 2}$ \\
$-$2400000 & km s$^{-1}$ & km s$^{-1}$      &  km s$^{-1}$ & km s$^{-1}$       \\
\hline
56672.3674 & 43.525      &  0.189           & 17.222       & 0.300 \\
57052.4032 & 43.272      &  0.189           & 16.272       & 0.301 \\
57053.3496 & 43.454      &  0.189           & 16.047       & 0.300 \\ 
57054.3817 & 43.694      &  0.188           & 15.918       & 0.300 \\
57055.3393 & 43.869      &  0.189           & 15.790       & 0.300 \\
57056.3651 & 43.881      &  0.189           & 15.774       & 0.300 \\
57237.7366 & 42.125      &  0.188           & 18.107       & 0.300 \\
57238.7311 & 41.513      &  0.189           & 18.983       & 0.301 \\
\hline
\multicolumn{5}{c}{HIP 20601} \\
 &&&&\\
 BJD         & $RV_1$      & $\sigma_{RV\;1}$ & $RV_2$       & $\sigma_{RV\; 2}$ \\
$-$2400000 & km s$^{-1}$ & km s$^{-1}$      &  km s$^{-1}$ & km s$^{-1}$       \\
\hline
55532.4785 &  25.8044 & 0.0423 &   62.9854 & 0.1795 \\
56243.5140 &  46.9337 & 0.0176 &   34.5268 & 0.2933 \\
56323.2404 & -24.2847 & 0.0196 &  130.7605 & 0.1366 \\
56323.3136 & -24.7068 & 0.0185 &  131.1920 & 0.1084 \\
56323.3628 & -24.9163 & 0.0198 &  131.5768 & 0.1179 \\
56323.4538 & -25.1587 & 0.0193 &  131.8981 & 0.1920 \\
56323.5102 & -25.2156 & 0.0224 &  132.0942 & 0.1792 \\
56324.2438 & -16.4490 & 0.0251 &  119.8076 & 0.1136 \\
56324.4318 & -12.0964 & 0.0125 &  114.1730 & 0.1463 \\
56324.4718 & -11.1097 & 0.0199 &  112.7543 & 0.2391 \\
56619.5265 &  33.7880 & 0.0315 &   52.4828 & 0.3178 \\
57009.4242 &  48.1646 & 0.0224 &   33.0725 & 0.2846 \\
\hline
\multicolumn{5}{c}{HIP 117186} \\
 &&&&\\
BJD         & $RV_1$      & $\sigma_{RV\;1}$ & $RV_2$       & $\sigma_{RV\; 2}$ \\
$-$2400000 & km s$^{-1}$ & km s$^{-1}$      &  km s$^{-1}$ & km s$^{-1}$       \\
\hline
55864.3650 &    -8.5826 & 0.8246 & -35.8419 & 0.2292  \\
56147.5270 &   -64.0310 & 0.8484 &  32.1576 & 0.1751  \\
56243.3282 &   -34.8861 & 1.4155 &  -3.0639 & 0.2020  \\
56525.5154 &     1.6961 & 0.4467 & -45.6927 & 0.1320  \\
56619.4355 &     2.7450 & 0.7522 & -47.3041 & 0.1390  \\                                                    
56889.5626 &    -2.5384 & 0.7092 & -41.1859 & 0.1439  \\                                                                      
56948.4278 &    -1.3476 & 0.7158 & -41.6893 & 0.1622  \\
\hline
\end{tabular}
\end{minipage}
\end{table}

The uncertainties of the RV measurements thus obtained need to be verified, and possibly
corrected, since the method of calculation leads to overestimating the errors. 
Again, the verification is based on $F_2$, and when a correction is necessary,
it is done in order to obtain $F_2=0$. For HIP~20601, we have 12 RV measurements for each component.
We derive the 6 parameters of each SB1 orbit, and $F_2$ is obtained with 6 degrees of freedom.
For the primary and for the secondary component, we obtain $F_2=-1.63$ and $-2.43$, respectively.
These small values clearly confirm that the uncertainties are too large for both components. They are multiplied
by 0.554 and 0.369, respectively, in order to have $F_2=0$. When the SB2 orbit is derived with the corrected
uncertainties, $F_2=0.32$. Although this value seems quite acceptable, we apply an additional correction factor 
of 1.058, in order to have $F_2=0$ for the SB2 orbit coming from our measurements, as we did for the SB2 orbit obtained
from previously published measurements, in section~\ref{sec:oldRV}.

We have only 7 RV measurements for each component of HIP~117186. This number is sufficient to derive the SB1 orbits,
but, with only one degree of freedom, the method applied to HIP~20601 is not sufficiently robust to lead to 
reliable uncertainties. Therefore, we consider only the SB2 orbit, which is derived with 6 degrees of freedom.
We have $F_2=-0.81$, and we correct the uncertainties by multiplying them by 0.766. After this correction, we verify
that both SB1 orbits have acceptable values of $F_2$: we find $F_2=-0.13$ and $-0.77$, respectively.

The RVs and the uncertainties finally derived are in Table~\ref{tab:VR}.

\subsection{The SB2 orbits}
\label{sec:orbSB2}

We compute the SB2 orbits, taking into account simultaneously the existing and the
new measurements. In addition to the usual parameters of an SB2 orbit, we introduce 3 offsets of the RV measurements: $d_{n-p}$, the offset between the new measurements and the published ones, and
$d^p_{2-1}$ and $d^n_{2-1}$, the offsets between the RV of the secondary components and the RV of the
primary components,
for the published and for the new measurements. 
The systemic velocity, $V_0$, is derived in the system of the new RV measurements of the primary component.

For HIP~14157 and HIP~20601, $F_2=-0.041$ and $-0.16$, respectively, indicating that both sets of
measurements are quite compatible.
For HIP~117186, $F_2=0.78$, since some discrepancies appear between the SB2 orbit derived from the new measurements and
the preceding one; the most important is the mass ratio, which is $0.771 \pm 0.021$ with the previously published measurements and
$0.844 \pm 0.012$ with our
observations. Nevertheless, the SB2 orbit obtained from both sets of RVs is basically indistinguishable from the one derived from our
measurements alone, but the period is much more accurate, thanks to the extension of the timespan covered
by the observations.
The new SB2 orbits are presented in Table~\ref{tab:orbSB2}.

\begin{table*}
 \centering
 \begin{minipage}{178mm}
  \caption{The orbital elements of the three stars, derived from both the previously existing RV measurements and from the
new ones. The minimum masses and minimum semi-major axes are derived from the true period ($P_{true}=P \times (1-V_0/c)$).}
\scriptsize
  \begin{tabular}{@{}lcccccccccccccc@{}}
  \hline
HIP & $P$ & $T_0$(BJD) & $e$ &  $V_0$ & $\omega_1$ & $K_1$ &  ${\cal M}_1 \sin^3 i$ &  $a_1 \sin i$&$N_1$& $d_{n-p}$ & $\sigma(O_1-C_1)_{p,n}$ \\
    &     &           &     &        &            & $K_2$ &  ${\cal M}_2 \sin^3 i$ &  $a_2 \sin i$&$N_2$& $d^p_{2-1},d^n_{2-1}$ &$\sigma(O_2-C_2)_{p,n}$    \\
    & (d) & 2400000+  &  &(km s$^{-1}$)&($^{\rm o}$)&(km s$^{-1}$)&(${\cal M}_\odot$)&  (Gm) & & (km s$^{-1}$) &(km s$^{-1}$)    \\
  \hline
14157  & 43.32058    & 51487.495 & 0.7602       & 30.751      & 174.60   & 54.31    &0.980     &21.014     &23+8& 0.343        &0.555,0.142\\
       &$\pm 0.00049$&$\pm 0.012$&$\pm 0.0015  $&$\pm  0.094 $&$\pm 0.22$&$\pm 0.29$ &$\pm0.010$&$\pm 0.094$&     & $\pm 0.176$   &         \\
       &             &            &              &            &           & 60.45    &0.8801     &23.39     &23+8& 0.434, -0.167  &0.636,0.312\\
       &             &            &              &            &           &$\pm 0.34$&$\pm0.0089$&$\pm 0.11$&     & $\pm 0.201$, $\pm 0.170$ & \\
&&&&&&&&&&&\\
20601  & 156.38019   & 56636.6716 & 0.85147      & 41.623     & 202.042   & 37.342    &0.9060     &42.101     &63+12& -0.442        &0.951,0.015\\
       &$\pm 0.00027$&$\pm 0.0027$&$\pm 0.00025 $&$\pm  0.014$&$\pm 0.089$&$\pm 0.017$&$\pm0.0037$&$\pm 0.025$&     & $\pm 0.112$   &         \\
       &             &            &              &            &           & 50.390    &0.6714     &56.81      &4+12 & 0.416, 0.071  &2.53,0.143\\
       &             &            &              &            &           &$\pm 0.088$&$\pm0.0018$&$\pm 0.10$ &     & $\pm 1.531$, $\pm0.136$ & \\
&&&&&&&&&&&\\
117186 & 85.8266     & 56403.36 & 0.3362      &-19.59    &178.75    &  33.40   &1.627     &37.12     &19+7&1.121          &2.14,0.799\\
       &$\pm 0.0017 $&$\pm 0.27$&$\pm 0.0035 $&$\pm 0.32$&$\pm 0.86$&$\pm 0.39$&$\pm0.028$&$\pm 0.44$&    & $\pm 0.527$   &         \\
       &             &          &             &          &          & 40.31    &1.348     &44.81     &19+7& 0.519, -0.967 &1.41,0.107\\
       &             &          &             &          &          &$\pm 0.18$&$\pm0.033$&$\pm 0.23$&    & $\pm 0.558$, $\pm 0.411$ &        \\
 \hline
\label{tab:orbSB2}
\end{tabular}
\end{minipage}
\end{table*}

%_________________________________________________________________________________________________________________

\section{The masses}
\label{sec:masses}

\subsection{Derivation of the masses}
\label{sec:Massederivation}

\begin{table}
 \centering
 \begin{minipage}{82mm}
 \caption{The combined VB+SB2 solutions; For consistency with the SB orbits and with the
forthcoming astrometric orbit, $\omega$ and $\Omega$ both refer to the motion of the primary
component.}
\label{tab:Masse}
\begin{tabular}{@{}lccc}
\hline
  &  HIP 14157 & HIP 20601 & HIP 117186 \\
\hline
$P$ (days)              &  43.32032   & 156.38020    & 85.8238          \\
                        &$\pm$ 0.00013& $\pm$ 0.00026& $\pm$ 0.0012     \\
$T_0$ (BJD-2400000)     & 51487.5005  & 56636.6713   & 56402.576        \\
                        &$\pm$ 0.0079 &$\pm$ 0.0027  & $\pm$ 0.072       \\
$e$                     & 0.7594      & 0.85148      & 0.32702         \\
                        &$\pm$ 0.0010 &$\pm$ 0.00025 & $\pm$ 0.00068   \\
$V_0$ (km s$^{-1}$)     & 30.743      & 41.623       & -19.89       \\
                        &$\pm$ 0.091  &$\pm$ 0.014   & $\pm$ 0.33    \\
$\omega_1$ ($^{\rm o}$) & 174.69      & 202.026      & 176.07 \\
                        &$\pm$ 0.17   & $\pm$ 0.086  & $\pm$ 0.32 \\
$\Omega_1$($^{\rm o}$; eq. 2000) &  19.141     & 340.526      & 16.928    \\
                        &$\pm$ 0.082  &$\pm$ 0.058   &$\pm$ 0.047 \\
$i$  ($^{\rm o}$)       & 92.24       & 103.138      & 88.054 \\
                        &$\pm$ 0.18   &$\pm$ 0.077   & $\pm$ 0.043 \\
$a$\footnote{the uncertainty refers to the VB solution} (mas) & 5.810 & 11.339       & 4.677 \\
                        & $\pm$ 0.034 &$\pm$ 0.068   & $\pm$ 0.032 \\
${\cal M}_1$ (${\cal M}_\odot$) & 0.982    & 0.9808  & 1.686  \\
                        &$\pm$ 0.010  & $\pm$ 0.0040 &$\pm$ 0.021 \\
${\cal M}_2$ (${\cal M}_\odot$) & 0.8819  & 0.7269   & 1.390  \\
                        &$\pm$ 0.0089 &$\pm$ 0.0019 &$\pm$ 0.034 \\
$\varpi$ (mas)          & 19.557      & 16.702      & 8.445  \\
                        &$\pm$ 0.078  &$\pm$ 0.037  &$\pm$ 0.075 \\
$n_{VLTI} \times 2$            & 20          & 12          & 14   \\
$\sigma_{(o-c)\;VLTI}$ (mas)  & 0.035 & 0.031      & 0.0084 \\
$n_{RV1}$               & 23+8        & 63+12       & 19+7 \\
$\sigma_{(o-c)\;RV1}$ (km s$^{-1}$)  & 0.562, 0.156 & 0.952, 0.015& 2.40, 0.89\\
$n_{RV2}$               & 23+8        & 4+12        & 19+7 \\
$\sigma_{(o-c)\;RV2}$ (km s$^{-1}$)  & 0.646, 0.282 & 2.54, 0.143 & 1.11, 0.23\\
$d_{n-p}$ (km s$^{-1}$)        & 0.323       & -0.441      & 0.822    \\
                               &$\pm$ 0.172  &$\pm$ 0.111  &$\pm$ 0.563 \\
$d^p_{2-1}$ (km s$^{-1}$)      & 0.408       & 0.436       & 0.549     \\
                               &$\pm$ 0.198  &$\pm$ 1.513  &$\pm$ 0.606 \\
$d^n_{2-1}$ (km s$^{-1}$)      & -0.149      & 0.077       & -0.184   \\
                               &$\pm$ 0.161  & $\pm$ 0.135 &$\pm$ 0.335 \\
\hline
\end{tabular}

\end{minipage}
\end{table}

The masses of the components are directly derived from the interferometric and from the RV
measurements, taken into account simultaneously. 
However, we increase the RV uncertainties of HIP~117186 by 1.088, which would lead to a SB2 orbit with $F_2=0$. This operation increases the
relative weights of the interferometric measurements in the derivation of the combined orbit.
The solution consists in up to 13 independent parameters,
which are: the orbital parameters $P$, $T_0$, $e$, $V_0$, $\omega_1$, $i$, $\Omega_1$, the masses ${\cal M}_1$,  ${\cal M}_2$,
the trigonometric parallax $\varpi$, and also the RV offsets $d_{n-p}$, $d^p_{2-1}$ and $d^n_{2-1}$.
It is worth noticing that we prefer to directly obtain ${\cal M}_1$,  ${\cal M}_2$ and $\varpi$,
rather than the observational parameters $K_1$, $K_2$ and $a$, the apparent semi-major axis of the
interferometric orbit. The advantage of this method is that it leads directly to the uncertainties of the
masses and of the parallax, in place of the uncertainties on $K_1$, $K_2$ and $a$ when the latter
parameters are obtained from
the combined interferometric and spectroscopic observations. The parameters of the 
combined solutions are presented in Table~\ref{tab:Masse}.
The uncertainties of masses range between 0.0019 and 0.034 ${\cal M}_\odot$, and the relative errors
range between 0.26 and 2.4~\%. This is similar to the accuracies expected using the Gaia astrometry.

\subsection{Notes on individual objects}
\label{sec:MasseNote}

\begin{description}

\item[\bf HIP 14157.] This system is extensively discussed by \cite{Fekel04}, who pointed out that the
primary is a BY Dra variable star with a variability amplitude around 0.02 mag. Due to an inclination
almost edge-on, the masses of the components are
close to the minimum masses that they found. We confirm then that the mass of the K2-K3~V
secondary component is 0.882~${\cal M}_\odot$. This is larger than the canonical value, which is between
0.67~${\cal M}_\odot$ (for a K5 V star), and 0.79~${\cal M}_\odot$ (for a K0~V star) according to
\cite{Allen}. Such discrepancy is not surprising, but known since a while, since \cite{Griffin85} and references therein
already pointed out that the real masses of K-type stars are usually 15~\% larger than the canonical values.
In a similar way, we find that 
the primary component is too heavy for a
K0 V star.
The minimum projected separation between the
components is only 0.090 mas, corresponding to 0.99 solar radius. Since Fekel et al. estimated the
stellar radii $R_1=0.99~{\cal R}_\odot$ and $R_2=0.76~{\cal R}_\odot$, the system is very likely an eclipsing one.

\item[\bf HIP 20601.] The star is a candidate member of the Hyades cluster \citep{Perryman98,deBruijne01}. \cite{Griffin85} estimated
that the spectral types of the components are probably G6 and K5, in good agreement with our estimates of the effective temperatures.
As a consequence, the masses are around
7~\% larger than the canonical values listed in \cite{Allen}.

\item[\bf HIP 117186.] 
The effective temperatures of the
components correspond to spectral types around F5,
and the canonical masses are around 1.4. This corresponds well to the mass of the secondary component,
but is around 20\% percent less than
the mass of the primary component.

\end{description}

%_________________________________________________________________________________________________________________

\section{The infrared mass-luminosity relation}
\label{sec:masse-lum}

\begin{table}
 \centering
 \begin{minipage}{70mm}
 \caption{The magnitudes in the infrared $H$ band, and the absolute magnitudes of the
components derived from the flux ratios taken from Table~\ref{tab:mesPIO} and from
the parallaxes in Table~\ref{tab:Masse}.}
\label{tab:H}
\begin{tabular}{@{}lccc}
\hline
  &  HIP 14157 & HIP 20601 & HIP 117186 \\
\hline
$H_{tot}$  & 6.629 $\pm$ 0.029 & 7.209 $\pm$ 0.047   & 6.252 $\pm$ 0.031 \\
$\Delta H$ & 0.429 $\pm$ 0.017 & 0.999 $\pm$ 0.016   & 0.891 $\pm$ 0.020 \\
$M_{H\;1}$ & 3.645 $\pm$ 0.031 & 3.687 $\pm$ 0.047   & 1.281 $\pm$ 0.037 \\
$M_{H\;2}$ & 4.073 $\pm$ 0.032 & 4.686 $\pm$ 0.049   & 2.172 $\pm$ 0.039 \\
\hline
\end{tabular}

\end{minipage}
\end{table}

The data derived from {\it PIONIER} observations include the flux ratios $f_2/f_1$ which are
listed in Table~\ref{tab:mesPIO}. The photometric band is similar to the infrared $H$
band of 2MASS \citep{2MASS}, and, for each binary, we derive the mean value of the magnitude
difference, $\Delta H$, and its standard error. The total $H$ magnitudes of the binaries are taken
from \cite{2MASS-PSC}, and the individual absolute $H$ magnitude of the components are then
computed, using the parallaxes from Table~\ref{tab:Masse}. The results are given in Table~\ref{tab:H}.

\begin{figure}
\includegraphics[clip=,height=2.2 in]{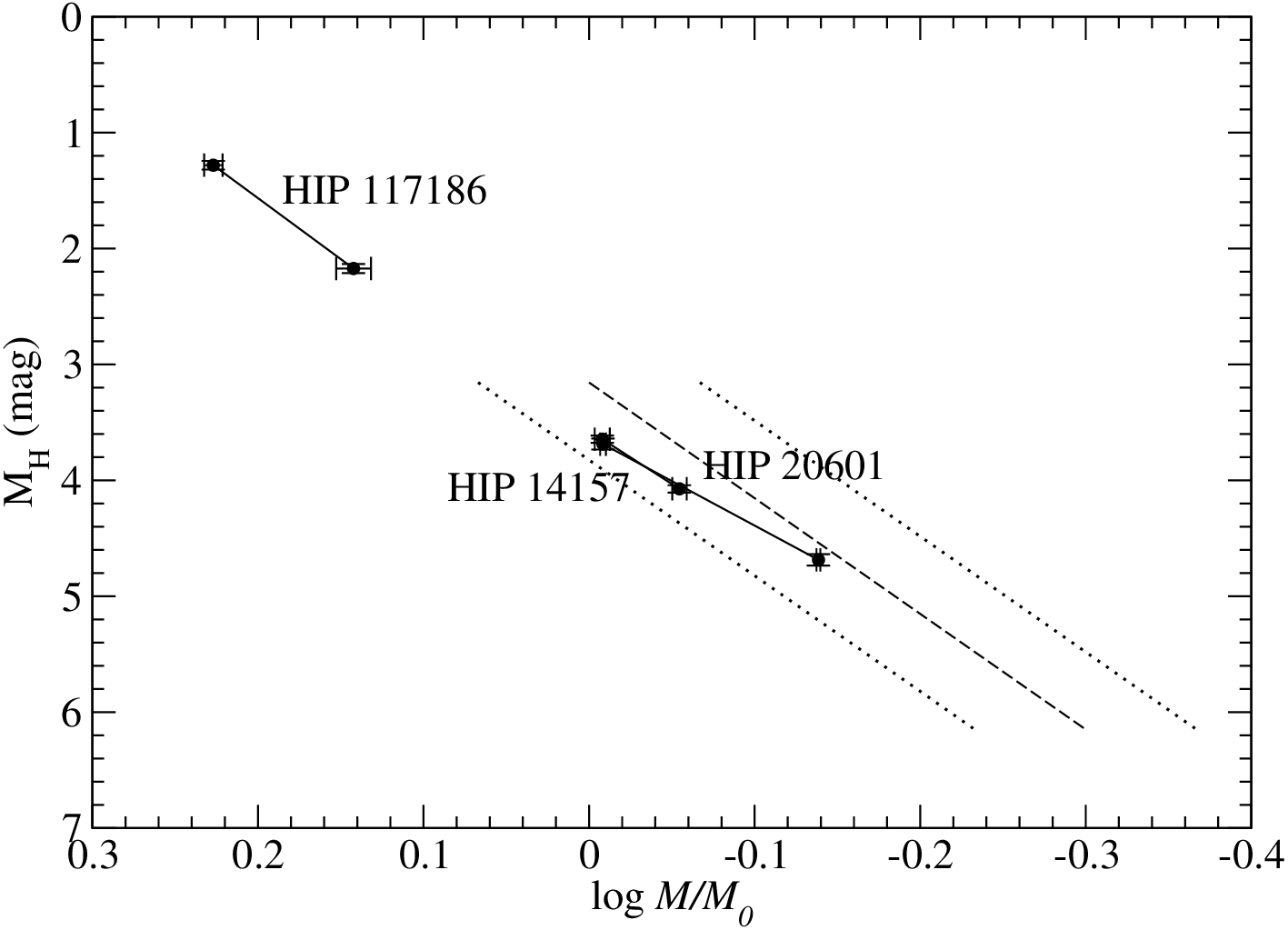}
 \caption{The mass-luminosity diagram in the infrared $H$ band. The
relation of Henry and McCarthy
is in dashes, with limits in dotted lines.
}
\label{fig:M-H}
\end{figure}

The masses and the absolute $H$ magnitudes of the six components are plotted on a
mass-luminosity diagram in Fig.~\ref{fig:M-H}. The mass-luminosity relation
of \cite{Henry93} is also shown for comparison. Although the masses of our stars are
within the uncertainties of the canonical relation, they are in excess by around 8~\%
when $H_{abs}$ is between 3.6 and 4.7 mag.

%_________________________________________________________________________________________________________________

\section{Verification of the {\it Hipparcos} parallaxes}
\label{sec:varpi}

\begin{table}
 \centering
 \begin{minipage}{70mm}
 \caption{The {\it Hipparcos 2} parallaxes, before and after taking into account the orbital
motion ($\varpi_{SS}$ and $\varpi_{AO}$, respectively). $a_0$ is the semi-major axis of the
photocentric orbit derived from {\it Hipparcos} data.}
\label{tab:HIP2}
\begin{tabular}{@{}lccc}
\hline
  &  HIP 14157 & HIP 20601 & HIP 117186 \\
\hline
$\varpi_{SS}$ (mas)        & 19.78 $\pm$ 1.10 & 15.20 $\pm$ 1.35 & 6.94 $\pm$ 0.57 \\
$F_{2\;SS}$              & 0.69             & -0.21            &       2.24      \\
$\varpi_{AO}$ (mas)        & 19.10 $\pm$ 1.09 & 14.84 $\pm$ 1.41 & 7.93 $\pm$ 0.64 \\
$a_0$ (mas)             & 3.06 $\pm$ 1.54  & 2.45 $\pm$ 1.40  & 1.07 $\pm$ 0.48 \\
$F_{2\;AO}$              & 0.399            & -0.42            &       2.03      \\
\hline
\end{tabular}

\end{minipage}
\end{table}

The elements of the combined solutions presented in Table~\ref{tab:Masse} include
parallaxes with errors between 0.037 and 0.078 mas. They are roughly 10 times better
than the errors of the parallaxes coming from {\it Hipparcos}. However, in the {\it Hipparcos 2}
catalogue, the parallaxes of these stars were derived through the single-star (SS) model, ignoring that they are binaries.
As a consequence, a discrepancy between our parallaxes and the {\it Hipparcos} ones could be due to
a reduction based on the use of the single--star model, and not to errors in the {\it Hipparcos}
transits. In order to check the reliability of {\it Hipparcos} itself, we first computed
the corrections of the {\it Hipparcos 2} parallaxes. For that purpose, the residuals of any single--star
solution were input in the computation of an astrometric orbital (AO) solution. However, except for
the astrometric semi-major axis, $a_0$, all the orbital elements were fixed on the values already
obtained. The new parallaxes, $\varpi_{AO}$, are listed in Table~\ref{tab:HIP2}, with $a_0$ and the
goodness-of-fit of the new solution, $F_{2\; AO}$. The uncorrected parallax and the related goodness-of-fit,
$F_{2\;SS}$, are indicated for comparison. $F_2$ is always ameliorated
when the orbital motion is taken into account, and $a_0$ is always smaller than twice its
uncertainty. Therefore, it would not have been not possible to detect the orbital motion from {\it Hipparcos} alone.

\begin{figure}
\includegraphics[clip=,height=2.2 in]{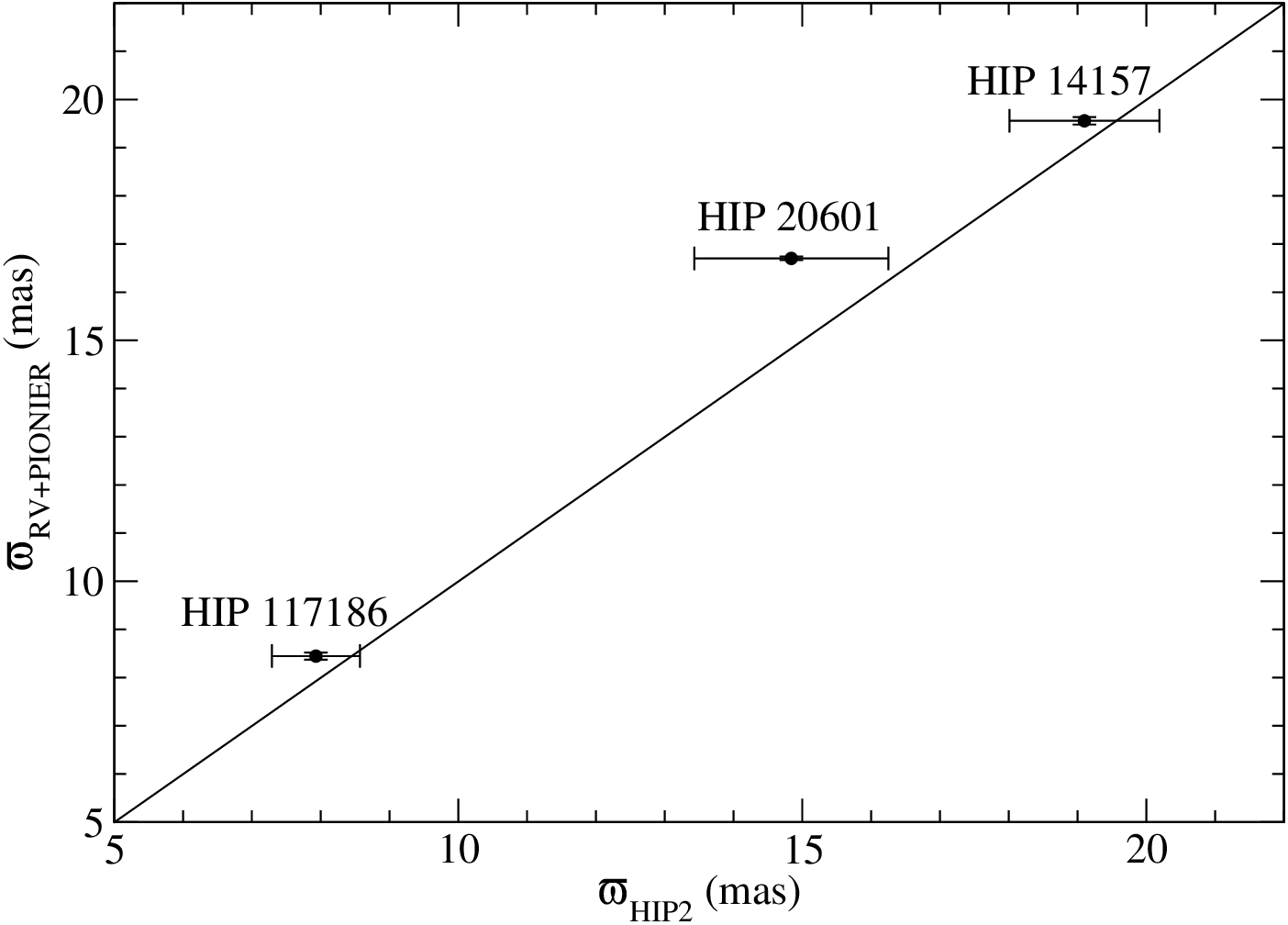}
 \caption{Comparison of the parallaxes derived in the combined solution with the
{Hipparcos 2} parallaxes corrected for the orbital motion.
}
\label{fig:varpi}
\end{figure}

The corrected {\it Hipparcos} parallaxes are compared to our parallaxes in Fig.~\ref{fig:varpi}.
For HIP~14157 and HIP~117186, the agreement is less than the standard error.
For HIP~20601, the difference is
($1.86 \pm 1.41$)~mas, ie 1.3 times the standard error. This is still a rather good agreement, and it seems
that the discrepancy found by Fekel for HD 207651 is due to peculiarities, such as the presence of a third star.

%_________________________________________________________________________________________________________________

\section{Conclusion}
\label{sec:conclusion}

We have combined interferometric observations performed with the {\it VLTI} with radial velocities in order
to derive the masses of the components of three binary stars. Thanks to the exquisite
accuracy of the {\it PIONIER} observations, but also to the fact that the orbits are all
close to edge-on, the accuracy of the masses thus obtained
is between 0.26 and 2.4 \%. This is less than the 3~\% limit applied by \cite{Torres10} when they
set up their list of accurate masses. This is also close to the uncertainties that we expect to
obtain combining the RV measurements with {\it Gaia} astrometry.

Five of the six masses are a few percent larger than the expectations
coming from the standard spectral type--mass calibration, confirming \cite{Griffin85}. The masses below
one solar mass are also around 8 \%
larger than the masses derived from the mass-luminosity
relation of \cite{Henry93}, although they are within their error interval. One of our star (HIP~14157)
should be observed as an eclipsing binary; this would confirm the small inclination that we have found,
and therefore improve the accuracy of the masses, but it would also make possible the estimation of
the radii of the components.

The parallaxes are derived in the same time as the orbital elements of the binaries, with an
accuracy much better than that of the {\it Hipparcos 2} catalogue. The reliability of the Hipparcos parallaxes
is confirmed.

%_________________________________________________________________________________________________________________

\section*{Acknowledgments}

This project was supported by the french INSU-CNRS ``Programme National de Physique Stellaire''
and ``Action Sp\'{e}cifique {\it Gaia}''.
{\it PIONIER} is funded by the Universit\'e Joseph Fourier (UJF), the Institut de Plan\'etologie et d'Astrophysique de Grenoble (IPAG),
and the Agence Nationale pour la Recherche (ANR-06-BLAN-0421, ANR-10-BLAN-0505, ANR-10-LABX56).
The integrated optics beam combiner is the result of a collaboration between IPAG and CEA-LETI based on CNES R\&T funding.
The {\it HERMES} spectrograph is supported by the Fund for Scientific Research of Flanders (FWO),
the Research Council of K.U.Leuven, the Fonds National de la Recherche Scientifique
(F.R.S.-FNRS), Belgium, the Royal Observatory of Belgium, the Observatoire de Gen\`eve,
Switzerland and the Th\"uringer Landessternwarte Tautenburg, Germany. 
We are grateful to the staff of the
Haute--Provence Observatory, and especially to Dr F. Bouchy, Dr H. Le Coroller, Dr M. V\'{e}ron, and the night assistants, for their
kind assistance. We warmly thank Dr. C. Soubiran for her helpful advices.
This research has received funding from the European Community's Seventh Framework Programme (FP7/2007-2013) under grant-agreement numbers 291352 (ERC).
This work made use of the Smithsonian/NASA Astrophysics Data System (ADS) and of the Centre de Donnees astronomiques de Strasbourg (CDS).

%$$$$$$$$$$$$$$$$$$$$$$$$$$$$$$$$$$$$$$$$$$$$$$$$$$$$$$$$$$$$$$$$$$$$$$$$$$$$$$$$$$$$$$$$$$$$$$$$$$$$$$$$$$$$$$$$$

\bsp

\label{lastpage}

\end{document}